# Isolated parkinsonism is an atypical presentation of *GRN* and *C9orf72* gene mutations

**Authors**

Fábio Carneiro, MD[1,2,3]#; Dario Saracino, MD[1,3,4]#; Vincent Huin, MD, PhD[5]; Fabienne Clot, PhD[6]; Cécile Delorme, MD[7]; Aurélie Méneret, MD, PhD[1,7]; Stéphane Thobois, MD, PhD[8]; Florence Cormier, MD, PhD[7]; Jean Christophe Corvol, MD, PhD[1,7,9]; Timothée Lenglet, MD, PhD[7,10]; Marie Vidailhet, MD, PhD[1,7]; Marie-Odile Habert, MD[11]; Audrey Gabelle, MD, PhD[12], Émilie Beaufils, MD[13,14]; Karl Mondon, MD, PhD[13]; Mélissa Tir, MD, PhD[15]; Daniela Andriuta, MD, PhD[15]; Alexis Brice, MD[1]; Vincent Deramecourt, MD, PhD[16]; Isabelle Le Ber, MD, PhD[1,3,7,17]*

# These authors equally contributed to this work.

**Affiliations**

1 - Sorbonne Université, Paris Brain Institute – Institut du Cerveau – ICM, Inserm U1127, CNRS UMR 7225, AP-HP - Hôpital Pitié-Salpêtrière, Paris, France.

2 - Hospital Garcia de Orta, Almada, Portugal.

3 - Centre de référence des démences rares ou précoces, IM2A, Département de Neurologie, AP-HP - Hôpital Pitié-Salpêtrière, Paris, France.

4 - Aramis Project Team, Inria Research Center of Paris, Paris, France.

5 - Univ. Lille, Inserm, CHU-Lille, Lille Neuroscience & Cognition, UMR-S1172, Team Alzheimer & Tauopathies, F-59000 Lille, France.

6 - Unité Fonctionelle de Neurogénétique Moléculaire et Cellulaire, Sorbonne Université, AP-HP, Hôpital Pitié-Salpêtrière, Paris, France.

7 - Département de Neurologie, AP-HP - Hôpital Pitié-Salpêtrière, Paris, France.

8 - Univ. Lyon, Université Claude Bernard Lyon 1, Faculté de Médecine Lyon Sud Charles Mérieux; CNRS, Institut des Sciences Cognitives Marc Jeannerod, UMR 5229, Bron; Hospices Civils de Lyon, Hôpital Neurologique Pierre Wertheimer, Neurologie C, Bron, France.

9 - Centre d'Investigation Clinique Neurosciences, AP-HP - Hôpital Pitié-Salpêtrière, Paris, France.

10 - Centre de Référence SLA-IdF, AP-HP - Hôpital Pitié Salpêtrière, Paris, France.

11 - Sorbonne Université, CNRS, Inserm, Laboratoire d'Imagerie Biomédicale, LIB, Paris; AP-HP - Hôpital Pitié-Salpêtrière, Médecine Nucléaire, Paris, France.

12 - CMRR, Département de Neurologie, CHU de Montpellier, Inserm U1061, Université de Montpellier i-site MUSE, Montpellier, France.

13 - Université François Rabelais de Tours, CHRU de Tours, Tours, France.

14 - Inserm U1253, IBrain, Tours, France.

15 - Département de Neurologie, Laboratoire de Neurosciences Fonctionnelles et Pathologies (UR UPJV 4559), Université d'Amiens et Université de Picardie Jules Verne, Amiens, France.

16 - Université de Lille, Inserm U1172, CHU Lille, DistAlz, LiCEND, CNR-MAJ, Lille, France.

17 - Paris Brain Institute – Institut du Cerveau – ICM, FrontLab, Paris, France.

**\* Corresponding author**

Dr Isabelle Le Ber

Paris Brain Institute – Institut du Cerveau – ICM.

Hôpital Pitié-Salpêtrière, 47-83, Boulevard de l'Hôpital.

75651 Paris Cedex 13, France.

Phone: 0033 1 57274682. Fax: 0033 1 5727 4795.

Email: isabelle.leber@upmc.fr

**Abstract**

**Introduction.** A phenotype of isolated parkinsonism mimicking Idiopathic Parkinson's Disease (IPD) is a rare clinical presentation of *GRN* and *C9orf72* mutations, the major genetic causes of frontotemporal dementia (FTD). It still remains controversial if this association is fortuitous or not, and which clinical clues could reliably suggest a genetic FTD etiology in IPD patients. This study aims to describe the clinical characteristics of FTD mutation carriers presenting with IPD phenotype, provide neuropathological evidence of the mutation's causality, and specifically address their "red flags" according to current IPD criteria.

**Methods.** Seven *GRN* and *C9orf72* carriers with isolated parkinsonism at onset, and three patients from the literature were included in this study. To allow better delineation of their phenotype, the presence of supportive, exclusion and "red flag" features from MDS criteria were analyzed for each case.

**Results.** Amongst the ten patients (5 *GRN*, 5 *C9orf72*), seven fulfilled probable IPD criteria during all the disease course, while behavioral/language or motoneuron dysfunctions occurred later in three. Disease duration was longer and dopa-responsiveness was more sustained in *C9orf72* than in *GRN* carriers. Subtle motor features, cognitive/behavioral changes, family history of dementia/ALS were suggestive clues for a genetic diagnosis. Importantly, neuropathological examination in one patient revealed typical TDP-43-inclusions without alpha-synucleinopathy, thus demonstrating the causal link between FTD mutations, TDP-43-pathology and PD phenotype.

**Conclusion.** We showed that, altogether, family history of early-onset dementia/ALS, the presence of cognitive/behavioral dysfunction and subtle motor characteristics are atypical features frequently present in the parkinsonian presentations of *GRN* and *C9orf72* mutations.

## 1. Introduction

The main clinical phenotypes of frontotemporal dementia (FTD) include behavioral variant of FTD (bvFTD), primary progressive aphasia (PPA), progressive supranuclear palsy (PSP) and corticobasal syndrome (CBS). They should not be regarded as discrete entities but rather as a continuum, with frequently overlapping clinical, pathological and genetic features. Parkinsonism is at the intersection of multiple syndromes within the FTD spectrum [1].

*C9orf72*, *GRN* and *MAPT* mutations are the leading genetic causes of FTD. The most frequently associated phenotype is bvFTD, with parkinsonian features often occurring during the disease course (FTD-parkinsonism) [2–5]. Less common clinical presentations are PSP, CBS and Lewy-body dementia (LBD)-like syndromes [3,5–11].

Isolated parkinsonian syndrome mimicking idiopathic Parkinson's Disease (IPD) much more rarely constitutes the only presenting feature of genetic FTD, in the absence of any cognitive/behavioral symptoms at onset. However, without pathological confirmation, the co-occurrence of sporadic Parkinson's disease (PD) in a carrier of a genetic FTD mutation cannot be excluded. Thus, it is currently unclear if this association is fortuitous or not, and which clinical clues, if any, could reliably suggest a genetic FTD etiology in patients with IPD.

The present study aims to: i) describe the clinical characteristics of patients presenting with isolated parkinsonism at onset associated with genetic FTD; ii) provide neuropathological evidence of the mutation's causality in this setting; iii) specifically address their "red flags" according to current IPD criteria.

## 2. Methods

### *2.1 Subjects*

Seven patients (3 *GRN*, 4 *C9orf72* carriers) with isolated parkinsonism mimicking IPD without cognitive/behavioral disorders at onset were identified as part of targeted next-generation sequencing of PD and FTD genes in a diagnosis or research setting. This study was approved by Paris-Necker ethics committee. All subjects signed informed consent. Clinical diagnoses of these patients and their relatives were reviewed according to the diagnostic criteria for PD [12], PSP [13], CBS [14], bvFTD [15].

*2.2. Genetic analyses*

All patients and/or their relatives underwent an extensive genetic screening of PD- and FTD-associated genes, including *GRN* and *C9orf72*, as detailed in Supplementary Methods [5,16]. Plasma progranulin levels were measured in *GRN* carriers (threshold: 71 ng/mL) [17].

*2.3. Pathological procedures*

Neuropathological examination was available for patient GRN-01 and performed on formalin-fixed, paraffin-embedded brain tissue samples. Paraffin sections were stained according to routine laboratory procedures. Immunohistochemistry (IHC) was achieved on a Ventana Benchmark automate, using commercially available antibodies directed against microtubule-associated protein Tau, alpha-synuclein, β-amyloid and TDP-43. Immunostaining was performed on samples from frontal, temporal and parietal associative cortices, anterior cingulate cortex, hippocampus, amygdala, putamen, midbrain, medulla oblongata, cervical spinal cord.

*2.4. Literature review*

A comprehensive literature review was performed to identify *GRN* or *C9orf72* carriers presenting with isolated parkinsonian symptoms at onset. A PubMed search used the terms: ((Frontotemporal dementia AND genetics) OR (Frontotemporal lobar degeneration AND genetics) OR (*GRN* OR *PGRN* OR progranulin) OR (*C9orf72* OR chromosome 9 open reading frame 72) AND (parkinson* OR progressive supranuclear palsy OR corticobasal syndrome OR corticobasal degeneration OR multiple system atrophy OR Lewy body) (Supplementary Table 1), from 2006 (year of *GRN* identification) to 2020.

A total of 527 articles were found, 61 of which were selected for full-text review, provided that the reported patients fulfilled the following criteria: i) carriers of *GRN* or *C9orf72* mutations with confirmed pathogenicity; ii) isolated parkinsonian syndrome at onset; iii) detailed description of the parkinsonian phenotype.

Fifty-five patients described in these papers met criteria for CBS (35 *GRN*, 6 *C9orf72*), PSP (2 *GRN*, 5 *C9orf72*) or LBD (4 *GRN*, 3 *C9orf72*) at onset and were not retained (Supplementary Table 2). Eleven patients had isolated PD at onset. For eight of them (mostly identified in large screening studies) the description was undetailed. Finally, two *GRN* and one *C9orf72* carriers with PD phenotype were included [18–20].

## 3. Results

### 3.1. Clinical phenotypes

Clinical descriptions of our seven cases are provided below. Pedigrees are displayed in Figure 1. The features of the overall series (our seven cases and three from the literature) are summarized in Table 1 and Supplementary Table 3. Some cases have been partially described elsewhere [10,16]. To allow better delineation of their phenotype, the presence of supportive, exclusion and "red flag" clinical features according to MDS criteria for PD were analyzed for each case (Supplementary Table 4, Supplementary Figure 1).

### 3.1.1. Patient GRN-01

Patient GRN-01 initially presented a rest tremor of the left hand at age 78, evolving to bilateral left-predominant and axial akinetic-rigid syndrome with camptocormia. Gait disturbances ensued two years later. L-dopa response was partial, although dose was limited to 150 mg by orthostatic hypotension (UPDRS-III on medication: 58). He had no oculomotor abnormalities. The clinical diagnosis was IPD.

Five years after onset, he progressively developed cognitive, notably executive dysfunction. No behavioral symptoms were reported. Mini-Mental State Examination (MMSE) was 20/30, and Frontal Assessment Battery (FAB) 8/18. Brain CT showed predominant frontotemporal atrophy. Five relatives had bvFTD or unspecified dementia. The patient and one sib affected by FTD carried a *GRN* deletion (exons 1-11). The patient died at age 86, eight years from onset. An autopsy was performed.

### 3.1.2. Patient GRN-02

Patient GRN-02 presented with generalized motor slowing predominating in the lower limbs at 53 years. He exhibited bilateral symmetric akinetic-rigid syndrome, hypomimia and axial rigidity without rest tremor (UPDRS-III on: 23). Eye movement recording was normal. Good L-dopa-responsiveness was observed. Dopamine transporter SPECT imaging (123I-FP-CIT) showed severe, predominantly right, pre-synaptic denervation (Figure 2). He was diagnosed with IPD. At age 56, he experienced severe motor fluctuations (wearing-off and unpredictable off).

Meanwhile, behavioral disturbances were absent. Global cognitive efficiency was preserved (MMSE 30/30), though moderate executive dysfunction was identified, with FAB scoring 15/18, Mattis Dementia Rating Scale (MDRS) 128/144 and Wisconsin card sorting test (WCST) 15/20. Brain MRI revealed mild bilateral prefrontal atrophy (Figure 2). At age 57, he presented visual (demons, animals) and kinesthetic hallucinations, and persecutory delusions occurring after titration of L-dopa and dopamine-agonists, only partially resolving with clozapine and after dopaminergic readjustment. At age 58, he showed mild eating behavior modifications (increased consumption of candies). Frontal cognitive syndrome worsened (MMSE 29/30; MDRS 119/144; FAB 13/18; WCST 3/20). Frontal atrophy was still moderate (Figure 2).

Although behavioral changes remained mild, he progressively became dependent on a walker and lost his autonomy for most daily-living activities. He died of viral pneumopathy at age 61, eight years after disease onset. The plasma progranulin level was decreased (25 ng/ml). The patient and one relative affected by CBS carried the *GRN* mutation p.Arg418* (p.R418X), a pathogenic mutation previously identified in FTD patients from other studies [21].

### 3.1.3. Patient GRN-03

Patient GRN-03 initially presented with a left upper limb postural myoclonic tremor at age 46. He rapidly developed an asymmetric akinetic-rigid parkinsonism, predominant in the left upper limb which exhibited good L-dopa response (UPDRS-III on: 10). EMG with back-averaging identified repetitive, 50-100 ms bursts with neither pre-myoclonic cortical potential nor cortical hyperexcitability. 123I-FP-CIT-scan revealed bilateral pre-synaptic striatal denervation. He received a clinical diagnosis of IPD. At age 48, he developed dysarthria and orofacial apraxia. Limb apraxia, cortical sensory deficit or alien-limb phenomenon were absent. Eye movement recording was normal.

At age 49, he progressively developed behavioral changes including apathy, binge eating, altered food preferences and loss of empathy. He scored 29/30 at the MMSE, 17/18 at the FAB, and 25/30 at the Montreal Cognitive Assessment (MoCA). Brain MRI and FDG-PET showed moderate temporal and mild prefrontal involvement, with sparing of thalamus and basal ganglia (Figure 2). At age 50, executive functions were altered. MMSE was 13/23, FAB 10/15, WCST 0/20. Ideomotor praxis were preserved. Shortly thereafter, the patient became mute and severely dysphagic. He died at age 51. Plasma progranulin level was 50 ng/ml. A *GRN* mutation, p.Glu498* (p.E498X), was identified in the patient. This mutation has already been described in one FTD patient in a previous study [22]. Five relatives had early dementia or PD.

*3.1.4. Patient C9-01*

The patient initially developed rest tremor of the left lower limb at age 48, sequentially spreading to both lower and upper limbs. At age 54, he presented bilateral lower-limb predominant bradykinesia, rigidity and rest tremor, with truncal anteflexion, gait festination and urinary urgency. 123I-FP-CIT-scan showed bilateral, right-predominant, striatal pre-synaptic denervation. He had good response to L-dopa and dopamine agonists (UPDRS-III off/on: 20/7). Two years later, motor fluctuations (wearing-off) and postural instability emerged. There was no oculomotor, autonomic, cognitive or cerebellar dysfunction. The clinical diagnosis was IPD.

At age 56, eight years after onset, mild attentional/executive difficulties were detected (digit span forward/reverse 3/2). At age 58, he developed apathy, loss of empathy and imitation behavior. MMSE score was 25/30, FAB 8/18, MDRS 118/144, and WCST 3/20.

There was progressive cognitive and motor deterioration, requiring wheelchair use at 59. MRI shortly before the patient's death revealed moderate frontal cortico-subcortical atrophy (Figure 2). The patient became bedridden and died at age 59. Three relatives had dementia, PSP or PD. The patient and his relative with PD carried a *C9orf72* expansion.

*3.1.5. Patient C9-02*

At age 29, patient C9-02 presented with left hemi-parkinsonism with akinesia, rigidity, rest tremor and dystonic posture. Motor involvement slowly became bilateral, although with sustained asymmetry, and good L-dopa response. At age 34, he had motor fluctuations and dyskinesias (UPDRS III off/on: 37/9; Hoehn and Yahr: 2). He met criteria for IPD. At age 37, cognitive functions were preserved. MMSE was 29/30, FAB 18/18, MDRS 138/144 and WCST 18/20. Brain MRI was normal. He received deep brain stimulation (DBS) of the subthalamic nucleus (STN) at age 37, with good therapeutic effect. Despite progression of motor symptoms, the patient maintains independence in daily-living activities at age 48, without evidence of cognitive/behavioral manifestations 19 years after disease onset. He carries a *C9orf72* expansion. Four family members had FTD, dementia, or ALS.

*3.1.6. Patient C9-03*

At age 64, the patient developed right akinetic-rigid parkinsonism (UPDRS-III on: 13) and right lower limb dystonia. L-dopa was effective. He received a clinical diagnosis of IPD. Motor fluctuations and dyskinesias ensued five years later, prompting apomorphine pump therapy at 69 years. He developed transient visual hallucinations (animals) under dopamine agonists.

At 69, global efficiency was preserved (MMSE 26/30) with a mild executive dysfunction (FAB 15/18) and no behavioral symptoms. Brain CT was normal. Motor and cognitive functions slowly deteriorated. MMSE was 20/30 at 73. At age 75, the patient was unable to undergo formal neuropsychological assessment (MoCA 10/30). He was wheelchair-bound at age 75 and lost to follow-up at 76, 12 years after onset. *C9orf72* expansion was detected. Two relatives had PD. Their DNA was not available.

*3.1.7. Patient C9-04*

At age 55, this patient presented a gait disorder with falls caused by bilateral tremor in lower limbs at rest and during walking. At age 57, he showed asymmetric, left-predominant akinesia and rigidity, axial rigidity and camptocormia partially responding to dopaminergic agents (UPDRS-III on: 27). He had a transient episode of paranoia, resolving after titration of dopaminergic treatment. REM sleep behavior disorder was reported. 123I-FP-CIT revealed bilateral striatal presynaptic denervation (Figure 2). He was diagnosed with atypical parkinsonism. Eye movement recording was normal. He had no dysautonomia.

At 58, he developed weakness in lower limbs and left upper limb, with amyotrophy, fasciculations, hyperreflexia and dropped head. Electromyography confirmed the diagnosis of ALS. Meanwhile, impulsiveness developed, without frank cognitive dysfunction. At age 59, MMSE was 29/30, FAB 18/18, MDRS 144/144 and WCST 20/20. MRI and FDG-PET revealed mild/moderate frontotemporal involvement (Figure 2). Progression was remarkably slow. The patient walked with a cane at 62 and required nocturnal non-invasive ventilation for respiratory failure at 69. There were no major behavioral, affective or emotional disorders at the last follow-up at age 70. He carried a *C9orf72* expansion. His mother developed early dementia. A sib had bipolar disorder.

*3.2. Summary of demographic and clinical features*

Among the overall series (n=10), including our cases and those previously described, the clinical course was compatible with a probable PD diagnosis according to MDS criteria in seven patients [12]. All the others had parkinsonian symptoms mimicking IPD, but two developed behavioral symptoms evocative of FTD (GRN-03) or ALS (C9-04) 3 years after onset and a third had early behavioral changes, aphasia and no response to L-dopa after 2 years [20]. No patients met criteria for other atypical parkinsonian syndromes, including PSP, CBS, MSA, and LBD.

The median age at onset was 54.0 years (range 29-78), with three patients showing early onset. Akinetic-rigid parkinsonism was present in all patients and was mostly asymmetric (n=8/9) and associated with rest tremor (n=7/10). Notably, lower-limb onset was frequent (6/9) including two presenting with hemi-parkinsonism. In all *C9orf72* carriers, lower limbs were the first affected site. Two patients exhibited postural myoclonic/jerky tremor, which was the first symptom for one of them (GRN-03).

L-dopa responsiveness was either good (n=6), partial (n=3) or absent (n=1), regardless of the genetic cause. One patient (C9-02) responded also to STN DBS. However, motor fluctuations and Dopa-induced dyskinesias tended to be more frequent in *C9orf72* than *GRN* carriers.

Executive dysfunction secondarily occurred in seven patients with a median time from disease onset of 4.0 years (IQR: 2.5; range: 2-8 years). Behavioral impairment occurred in six patients at a median of 3.0 years from onset (IQR: 1.9; range: 2-10 years) and was present in most *GRN* (4/5) and few *C9orf72* carriers (2/5).

The relatively short disease duration (median disease duration at death: 7.0 years; IQR: 3.0; range: 5-11 years) in the cohort was mostly driven by *GRN* carriers, who had more rapid progression (Supplementary Table 3). Though groups were too small for statistical comparisons, *C9orf72* carriers had longer median disease duration reaching 13.0 years (IQR 3.5) at the last follow-up.

All patients had family history of dementia and/or ALS except one in our series (C9-03), who only had family history of PD, and one patient in the literature who had no family history of neurodegenerative disorders [20].

### *3.3. Neuropathology*

Patient GRN-01 underwent neuropathological examination (Figure 3). The left cerebral hemisphere weighed 388g, with moderate cortical atrophy. Basal ganglia were structurally intact.

The frontal and temporal lobes were most affected, nonetheless with mild laminar spongiosis of cortical layers I-III. There was marked neuronal loss of the substantia nigra (SN). IHC revealed neuronal cytoplasmic and intranuclear TDP-43-positive inclusions and rare dystrophic neurites. TDP-43 pathology was most marked in the hippocampus and temporal lobes and was remarkably absent from the striatum and the midbrain. Alpha-synuclein IHC was negative all over the brain, including the cortex and the SN. Beta-amyloid pathology was absent. Neurofibrillary tangles were present in the entorhinal cortex (Braak stage I). There were no astrocytic plaques or tufted astrocytes. The pathological diagnosis was FTLD-TDP type A. There were no pathological arguments for IPD.

### 4. Discussion

Parkinsonism is a frequent feature in sporadic FTD, most commonly presenting after onset of bvFTD, as symmetric akinetic-rigid syndrome with limited L-dopa response

[23,24]. A similar parkinsonian phenotype is associated with behavioral changes in *C9orf72* [4,25] and *GRN* carriers [5,26,27].

Our findings illustrate that, although rare, an initial phenotype mimicking IPD can be associated with *GRN* and *C9orf72* mutations. In our series, 70% presented with rest tremor, most exhibiting asymmetric signs and satisfying response to L-dopa. However, some unusual features were observed. Postural myoclonic/jerky tremor, in isolation or associated with rest tremor, was a remarkable feature in two *GRN* carriers. Accordingly, postural tremor has been previously underlined in *GRN*-associated parkinsonism [28]. The distinctive features in *C9orf72* patients were the lower-limb onset, in all, and the favorable L-dopa responsiveness, often accompanied by dyskinesias, resembling IPD and differing from *GRN* carriers. Notably, a meta-analysis of parkinsonism in genetic FTD revealed that L-dopa was more frequently ineffective in *GRN* (75.5%) than *C9orf72* carriers (21.5%) [3].

Cognitive dysfunction was more frequent in our series (80%) compared to IPD cohorts (26%) [29] and occurred earlier [30]; the appearance of behavioral changes other than apathy (disinhibition, hyperorality, dietary changes) was remarkable. Frontal involvement on brain imaging was also distinctive, even if relatively modest for *GRN* and *C9orf72* carriers. These characteristics may serve as red flags to guide clinicians to appropriately test FTD genes.

Moreover, our study shows that a family history of dementia (particularly early-onset) or ALS should be carefully searched. It was present in eight of our patients, thus providing useful information to look for FTD mutations in patients with IPD phenotype. However, family history was absent in one patient and positive only for PD in another. Therefore, even without compatible family history, *C9orf72* and *GRN* should be analyzed if additional clues stemming from accurate motor, cognitive and behavioral evaluations are present. Additionally, the current availability of plasma progranulin dosage may provide a useful and cost-effective biomarker to easily identify potential *GRN* carriers [17]. Unfortunately, metaiodobenzylguanidine (MIBG) cardiac scintigraphy was not available for any of our patients. Its contribution would be valuable in these cases as it could potentially provide additional clues to differentiate parkinsonian syndromes related to FTD gene mutations from IPD.

Although the frequency of FTD mutations in IPD cohorts is an important question, we could not accurately determine it in the present study because our patients were identified through heterogeneous cohorts, and genetic analyses were performed either in a research setting or for diagnostic purposes. Therefore, clinico-genetic studies with appropriate design are needed to address this relevant question. However, previous genetic screenings have already suggested the rarity of FTD mutations in large PD cohorts [18,31]. In an extensive literature review, we identified only three well-described cases, enlarging our cohort. As IPD is a frequent neurodegenerative disease, it is questionable whether these cases represent a coincidental association of sporadic IPD and FTD gene mutations. Without neuropathological confirmation, we cannot exclude the co-occurrence of alpha-synucleinopathy in most reported cases. Rare *post-mortem* descriptions of *GRN* [18,26] or *C9orf72* carriers [32] with PD phenotype indeed revealed combined TDP-43 and diffuse Lewy-body pathologies affecting neocortical and subcortical areas, including the SN. Our study provided clear evidence that some FTD genes causing TDP-43 pathology may occasionally underlie the IPD phenotype, without concomitant alpha-synucleinopathy.

Clinical phenotypes in neurodegenerative diseases largely depend on the lesional topography. As such, different pathologies affecting the same regions cause similar syndromes. In GRN-01, degeneration of the SN provides a pathophysiological basis for his presentation. This is concordant with the finding that *GRN* patients had greater SN pathology than *GRN*-negative FTD-TDP patients [28], and explains the pre-synaptic denervation on dopamine transporter imaging. In FTD, basal ganglia are affected in late stages, after frontal and temporal cortices [33], and parkinsonian features usually occur during the disease course [23,24]. Our work suggests that, in some genetic cases, the pathological process may preferentially involve the SN and basal ganglia in the initial stages, resulting in IPD-like phenotype. Some genetic factors might explain those different patterns, as suggested by the familial aggregation of parkinsonian phenotypes in C9-01 and C9-03, and in some reported *GRN* and *C9orf72* families [26,34].

In conclusion, this study showed that: 1) *GRN* and *C9orf72* mutations may present with isolated parkinsonism fulfilling criteria for probable IPD; 2) those parkinsonian syndromes may occur independently of a coincidental alpha-synucleinopathy, pointing to a causal role of FTD mutations; 3) the most consistent clues to evoke FTD causative mutations, though not universally present, include family history of early-onset dementia/ALS, early or severe cognitive dysfunction, and behavioral changes. More broadly, these findings might be valuable for clinical practice in movement disorders. Noteworthy, family history of dementia and ALS should be potentially appraised as additional red flags in IPD criteria.

**Acknowledgements**

We thank the DNA and cell bank of the ICM (ICM, Paris), Kathy Larcher (UF de Neurogénétique, Pitié-Salpêtrière Hospital, Paris), Sandrine Noël (UF de Neurogénétique, Pitié-Salpêtrière Hospital, Paris), and Isabelle David (UF de Neurogénétique, Pitié-Salpêtrière Hospital, Paris) for their technical assistance.

**Funding**

The research leading to these results received funding from the "Investissements d'avenir" ANR-11-INBS-0011. This work was funded by the Programme Hospitalier de Recherche Clinique (PHRC) FTLD-exome (to ILB, promotion by Assistance Publique – Hôpitaux de Paris) and by PHRC Predict-PGRN (to ILB, promotion by Assistance Publique – Hôpitaux de Paris)

**Disclosures**

CD received travel grants from Merz Pharma, Medtronic and Boston Scientific. AM received travel grants from Merz Pharma and Abbvie. ST received research grants from ANR, Neurodis, France Parkinson, FRM; honoraria from Boston, Aguettant, Novartis, Teva; travel grant and congress registration grants from Abbvie, Zambon, Ellivie. JCC served as a member of advisory boards for UCB, Biogen, Prevail Therapeutic, Idorsia, Sanofi, Ever Pharma, Denali, BrainEver, Theranexus, and received grant from the Michael J Fox Foundation outside the present work. MOH has received fees as a consultant from Blue Earth company. MT has received travel grants from Abbvie, Orkyn, Medtronic, Boston, Elivie, Adelia, outside the submitted work. DA has received travel grants from Biogen, Roche, Teva, Novartis, Bristol-Myers Squibb, Genzyme, Sanofi, outside the submitted work. ILB served as a member of advisory boards for Prevail Therapeutic and received research grants from ANR, DGOS, PHRC, Vaincre Alzheimer Association, ARSla Association, Fondation Plan Alzheimer outside of the present work.

**Authors' roles**

All authors made substantial contributions to the work and approved the final article. FC, DS: conceptualization, acquisition and analysis of data, writing the original draft. VH,

**Table 1. Clinical characteristics of *GRN* (n=5) and *C9orf72* (n=5) patients in the present series and from the literature.**

| Gene | *GRN* | | | | | *C9orf72* | | | | |
|---|---|---|---|---|---|---|---|---|---|---|
| Study | Brouwers 2007 | Carecchio 2014 | This study | | | Annan 2013 | This study | | | |
| Subject | DR205.1 | Proband | GRN-01 | GRN-02 | GRN-03 | Proband | C9-01 | C9-02 | C9-03 | C9-04 |
| Age of onset (years) | 55 | 51 | 78 | 53 | 46 | 63 | 48 | 29 | 64 | 55 |
| DD (years) | 6† | 3* | 8† | 8†** | 5† | 3* | 11† | 19* | 12* | 15* |
| Diagnosis at onset | Probable PD | Probable PD | Probable PD | Probable PD | Probable PD | Probable PD | Probable PD | Probable PD | Probable PD | Atypical Park |
| Prodromal features | NA | NA | NA | Constipation, Erectile dysfonction | Constipation | NA | Constipation Sleep disorders Urinary urgency | - | NA | Depression, RLS, RBD |
| First symptoms | R, A, T | T | T | A, R | Myocl. T | A, T | T | A, T | R, A | T |
| Parkinsonism | | | | | | | | | | |
|   Criteria | R, A, T | R, A, T | R, A, T | R, A | R, A | R, A, T | R, A, T | R, A, T | R, A | R, A, T |
|   Site at onset | NA | Upper L | Upper L | Lower L | Upper L | Up/Lo L | Lower L | Up/Lo L | Lower L | Lower L |
|   Symmetry | NA | Asym | Asym | Sym | Asym | Asym | Asym | Asym | Asym | Asym |
|   Gait impairment | + | + | + | - | - | - | + | - | + | + |
|   Postural instability | + | NA | + | - | - | NA | + | - | - | + |
|   UPDRS-III on/off | NA | NA | 58/NA | 23/NA | 10/NA | NA | 7/20 | 9/37 | 13/NA | 27/NA |
|   Dopa-responsiveness | + | - | + (partial) | + | + | + (partial) | + | + | + | + (partial) |
|   MF / Dyskinesias | NA/NA | NA/NA | -/- | +/- | -/- | -/- | +/- | +/+ | +/+ | -/- |
| L-dopa induced hall. | NA | - | - | + | - | - | - | - | + | - |
| Cognitive impairment | + | + | + | + | + | + | + | - | + | - |
|   Years since onset | 2 | 2 | 5 | 3 | 4 | 2 | 8 | | 5 | |
| Behavioral changes | + | + | - | (+) | + | - | + | - | - | + |
|   Years since onset | 2 | 2.5 | | 5 | 3 | | 10 | | | 3 |

| Other signs/symptoms | - | Rest and postural jerky tremor | Camptocormia | Delusions | Orofacial apraxia | - | - | Dystonia | Dystonia | Camptocormia Delusions |
|---|---|---|---|---|---|---|---|---|---|---|
| Secondary diagnosis (years from onset) | PD + frontal dysfunction (2) | PD + Probable FTD (2) | Probable PD | Probable PD | PD + Probable FTD (3) | Probable PD | Probable PD | Probable PD | Probable PD | PD + ALS (3) |
| Family history | Dementia | No family history | Dementia, FTD | Dementia, CBS | Dementia, PD | Dementia, FTD | PD, PSP, Dementia | Dementia, FTD, ALS | PD | Dementia, BD |

+: present; (+): mild; -: absent; †: deceased; *: at last follow-up; ** cause of death: viral pneumopathy; A: akinesia; ALS: amyotrophic lateral sclerosis; Asym: asymmetric; BD: bipolar disorder; CBS: corticobasal syndrome; DD: disease duration; FTD: frontotemporal dementia; Hall.: hallucinations; L: limb; Lo: lower; MF: motor fluctuations; Myocl. T: myoclonic tremor; NA: not available; Park: parkinsonism; PD: Parkinson's disease; PSP: progressive supranuclear palsy; R: rigidity; RBD: REM-sleep behavioral disorder; RLS: restless legs syndrome; Sym: symmetric; T: tremor; Up: upper; UPDRS-III: Unified Parkinson's Disease Rating Scale – Part III.

# Figures

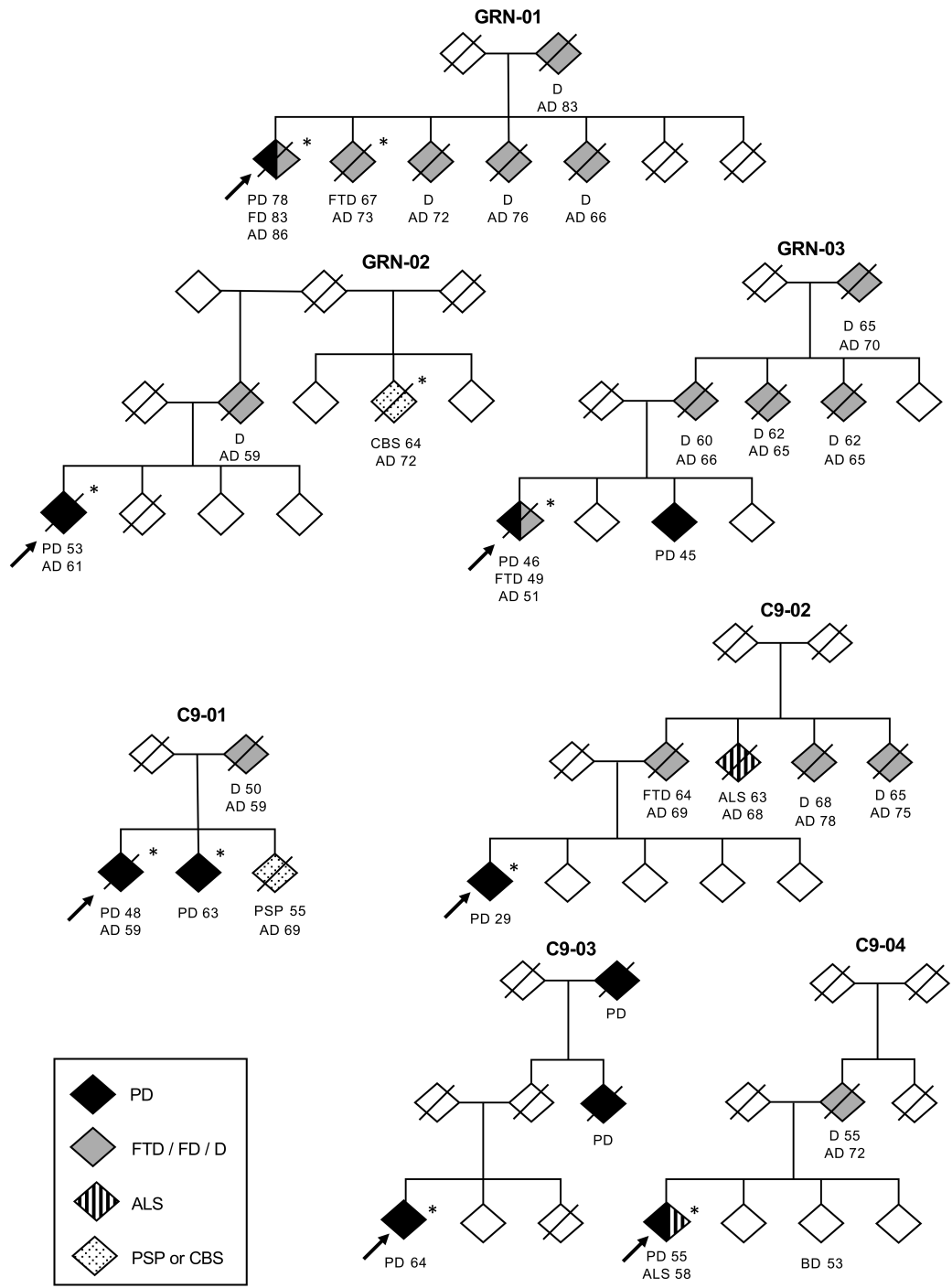

**Figure 1. Pedigrees of patients from our series.** Age at onset is provided after each diagnosis. *: patients carrying *GRN* mutation/*C9orf72* expansion; AD: age at death; ALS:

amyotrophic lateral sclerosis; BD: bipolar disorder; CBS: corticobasal syndrome; D: dementia (unspecified); FD: frontal dysfunction; FTD: frontotemporal dementia; PD: Parkinson's disease; PSP: progressive supranuclear palsy. Arrows indicate the patients described in this study. For confidentiality, pedigrees were slightly modified and genders were masked (diamonds).

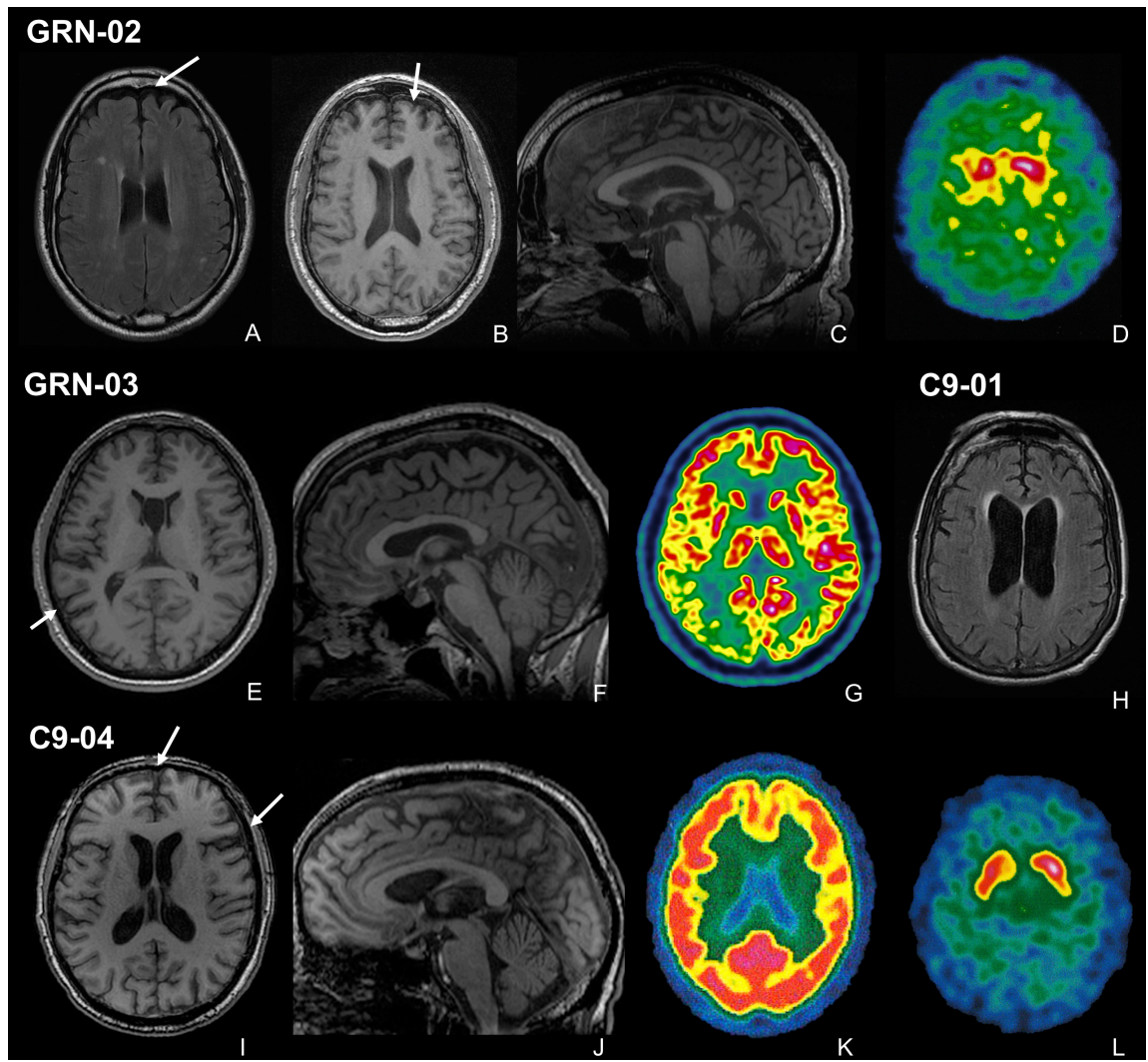

**Figure 2. Structural and functional neuroimaging in patients GRN-02, GRN-03, C9-01 and C9-04.** In patient GRN-02, brain MRI three years after disease onset (A) showed small white matter hypersignals and mild bilateral prefrontal atrophy (arrow). Five years after onset (B, C) frontal atrophy was still moderate. There was no midbrain or pontine atrophy. Dopamine transporter SPECT imaging (123I-FP-CIT) showed severe bilateral presynaptic denervation, predominantly on the right (D). Patient GRN-03 presented moderate temporal and mild prefrontal atrophy at three years from onset (E, F). FDG-PET showed asymmetric right-predominant hypometabolism of temporal cortex, with sparing of thalamus and basal ganglia (G). In patient C9-01, the MRI eleven years after disease onset showed moderate cortico-subcortical atrophy with frontal predominance

(H). Patient C9-04 had mild/moderate frontotemporal atrophy (arrows) without brainstem or cerebellar atrophy, three years after onset (I, J). FDG-PET disclosed discrete frontotemporal hypometabolism (K). Dopamine transporter SPECT imaging (123I-FP-CIT) showed bilateral presynaptic denervation (L).

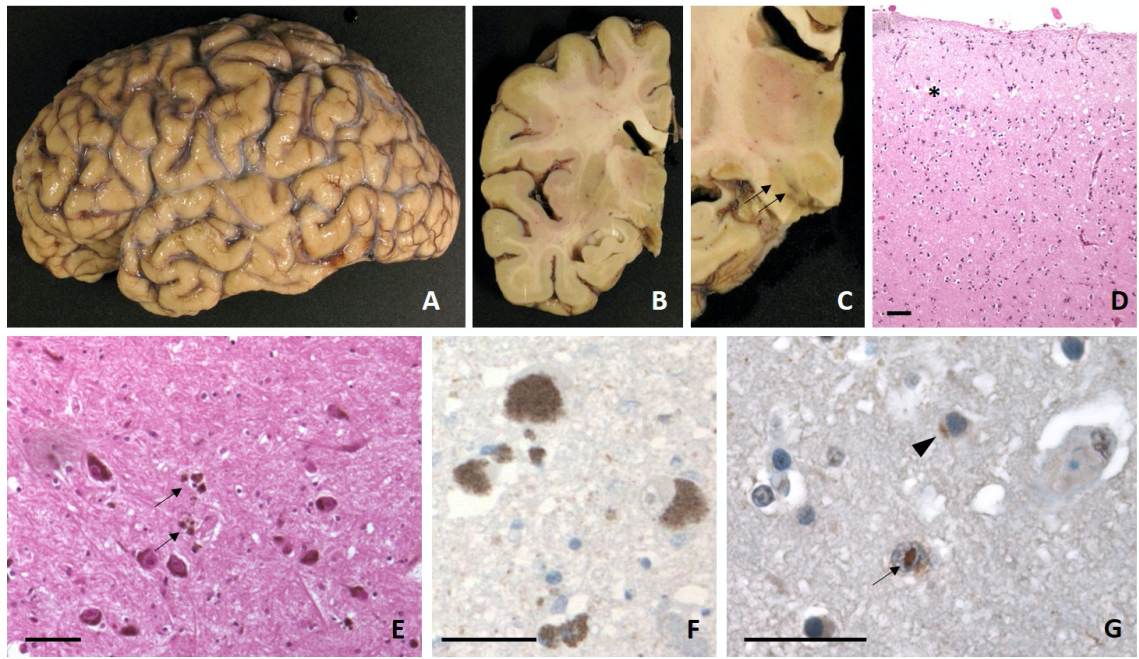

**Figure 3. Neuropathologic examination of patient GRN-01.** A: Lateral view of the left fixed hemisphere showing moderate cortical frontotemporal atrophy. B: Frontal cut through the left fixed hemisphere at the level of the mammillary body showing globally preserved basal ganglia appearance. C: Close-up view of the midbrain region showing severe pallor of the substantia nigra (arrows). D: Microscopic view of the frontal cortex after hematoxylin-eosin staining showing mild neuronal loss and laminar spongiosis of the upper cortical layers (asterisk). E: Microscopic view of the substantia nigra after hematoxylin-eosin staining showing significant loss of dopaminergic neurons with extracellular pigmented remnants (arrows) but no specific neuronal inclusion. F: Pigmented dopaminergic neurons in the substantia nigra after alpha-synuclein immunohistochemistry showing the absence of Lewy bodies. G: Frontal cortex neurons after TDP-43 immunohistochemistry showing an intranuclear neuronal lentiform inclusion (arrow) and a somatic neuronal inclusion (arrowhead) consistent with the diagnosis of FTLD-TDP pathology. Scale bar for all microscopic images: 100μ.